\documentclass[hyper]{JHEP} 

\usepackage{epsfig}

\newcommand\fverb{\setbox\pippobox=\hbox\bgroup\verb}
\newcommand\fverbdo{\egroup\medskip\noindent%
			\fbox{\unhbox\pippobox}\ }
\newcommand\fverbit{\egroup\item[\fbox{\unhbox\pippobox}]}
\newbox\pippobox
\title{Note About Tachyon Kink In 
Nontrivial Background}

\author{by J. Kluso\v{n}\\
     Department of Theoretical Physics and Astrophysics\\

                   Faculty of Science, Masaryk University\\

Kotl\'{a}\v{r}sk\'{a} 2, 611 37, Brno\\

Czech Republic\\

    E-mail: \email{klu@physics.muni.cz}}

\preprint{\hepth{0506250}}

\abstract{This paper is devoted to the
study of the tachyon kink on 
the worldvolume
of a  non-BPS Dp-brane that moves in
a nontrivial background. 
We will show that the spatial
dependent  tachyon condensation
leads to an emergence of a D(p-1)-brane whose
 dynamics
is governed by Dirac-Born-Infeld action.}

\def\bA{\mathbf{A}}

\def\bAi{(\mathbf{A}^{-1})}
\def\mat{\tilde{\mathbf{a}}}

\def\mati{(\tilde{\mathbf{a}}^{-1})}

\begin{document}
\section{Introduction}\label{first}
Study of various aspects of the tachyon
dynamics on a non-BPS Dp-brane in
type IIA or IIB theories has led to
some understanding of the tachyon
dynamics near the tachyon vacuum
\footnote{For review of the open string
tachyon condensation, see 
\cite{Sen:1999mg,Ohmori:2001am,
Taylor:2002uv,
Taylor:2003gn,Sen:2004nf}.}. The 
tachyon effective action (\ref{acg}), describing
the dynamics of the tachyon field
on a non-BPS Dp-brane of type IIA and
IIB theory was proposed in 
\cite{Sen:1999md,Bergshoeff:2000dq,
Garousi:2000tr,Kluson:2000iy}. It
was argued in many papers that
 the tachyon 
effective action (\ref{acg}) 
gives  a good description
of the system under condition that
tachyon is large and the second
and higher derivatives of the tachyon
are small
\footnote{For discussion of the
effective field theory description of
the tachyon condensation, see
\cite{Sen:2002qa,Fotopoulos:2003yt,
Sen:2003bc,Kutasov:2003er,
Niarchos:2004rw}.}. A kink solution
in the full tachyon effective field
theory, which is  supposed to describe
a BPS D(p-1)-brane was also constructed
in  
\cite{Sen:2003tm,Kim:2005he,Kim:2003in,
Kim:2003ma,Banerjee:2004cw,Bazeia:2004vc,
Copeland:2003df,Brax:2003rs}. A kink solution,
that by definition interpolates between
the vacuua at $T=\pm \infty$ has to
pass through $0$. Then we could expect that
higher derivative corrections to the
tachyon  effective action will be needed
to provide a good description of
the D(p-1)-brane as a kink solution. 
This issue was carefully analysed in
 paper \cite{Sen:2003tm} where it
was shown that the energy density 
of the kink in the effective field
theory is localised on codimension
one surface as in the case of
a BPS D(p-1)-brane. It was then also
shown that the worldvolume theory
of the kink solution is also given
by the Dirac-Born-Infeld (DBI) action
on a BPS D(p-1)-brane. Thus result
shows that the kink solution of the 
effective field theory does provide
a good description of the D(p-1)-brane
even without taking into account higher
derivative corrections. In other
words, the tachyon effective action
reproduces the low energy effective
action on the world-volume of the
soliton without any correction terms.

Since these results  are very
impressive  it would be certainly
useful to test 
the  effective field theory description
of the tachyon condensation 
in other, more general situations.
In fact, since the DBI action 
 describes the low
energy dynamics of the BPS Dp-brane 
in general curved background we
can ask the question whether we
can construct the tachyon kink on the
worldvolume of a non-BPS Dp-brane
embedded in curved background and whether
this kink has the interpretation as
a lower dimensional D(p-1)-brane
\footnote{Some partial results considering
tachyon condensation on non-BPS Dp-brane
in curved background were presented in
\cite{Kim:2005he,Panigrahi:2004qr,Kluson:2004yk,
Kluson:2005qx}.}.
To answer this question
  we begin 
 with common presumption that
 the tachyon effective action 
for Dp-brane (\ref{acg}) 
can be applied for the description of the tachyon
dynamics in the nontrivial background
 \footnote{The case of
more general background, including 
NS
$B$ field and Ramond-Ramond forms
 will be discussed in
forthcoming publication.}as well.
Then we will study the equation
of motion for the tachyon and
for the modes that parametrise
the embedding of the unstable
Dp-brane in given spacetime.
We will solve these equations
with the field configuration similar
to the ansatz that was given 
 in  \cite{Sen:2003tm}.
We will show that  this ansatz solves
the equation of motion for tachyon
on condition that the mode 
$t$ that characterises the core
of the kink (The precise meaning
of this claim will be given bellow.)
satisfies the equation of motion
of the scalar field 
 that describes the
embedding of D(p-1)-brane in given
background. This result  shows
that the spatial dependent tachyon
condensation leads to the emergence
of a D(p-1)-brane where the scalar
modes that propagate on its worldvolume
 solve the equation of
motion that arise from the DBI action
for D(p-1)-brane that is moving
in  the same  background. 

The structure of this paper is
as follows. In the next section
(\ref{second}) we will analyse 
the equation of motion for 
non-BPS Dp-brane in curved background.
We will find the spatial dependent
tachyon solution that has interpretation
as a lower dimensional D(p-1)-brane
whose dynamics is governed by DBI 
action. 
In section (\ref{third}) we will
study some  examples of 
the nontrivial background. The first
one corresponds to the stack
of $N$ NS5-branes  and the
second one corresponds to the
background generated by the collection
of $N$ coincident Dk-branes. 
Finally, in conclusion (\ref{fourth})
we will 
outline our results and suggest
possible extension of this work.

\section{Non-BPS Dp-brane
in general background}\label{second}
The starting point for the analysis
of the dynamics of a 
non-BPS Dp-brane
in general background
is the Dirac-Born-Infeld like
tachyon effective action 
\cite{Sen:1999md,Bergshoeff:2000dq,
Garousi:2000tr,Kluson:2000iy}
\footnote{We use the convention
where the fundamental string tension
has been set equal to $(2\pi)^{-1}$
(i.e. $\alpha'=1$).}
\begin{eqnarray}\label{acg}
S=-\int d^{p+1}\xi
e^{-\Phi}V(T)\sqrt{-\det \bA} \ ,
\nonumber \\
\bA_{\mu\nu}=g_{MN}
\partial_\mu Y^M\partial_\nu Y^N+
 F_{\mu\nu}+
\partial_\mu T\partial_\nu T \ , 
\mu \ , \nu=0,\dots, p \ ,
\nonumber \\
F_{\mu\nu}=\partial_\mu A_\nu
-\partial_\nu A_\mu \ ,
\nonumber \\
\end{eqnarray}
where $A_\mu \ , 
\mu,\nu=0,\dots,p$ and $ Y^{M,N} \ , 
M,N=0,\dots,9$ are gauge and the
transverse scalar
fields on the worldvolume of 
the non-BPS Dp-brane and 
$T$ is the
tachyon field.
Since in this paper 
 we will restrict ourselves
to  the situations when
the gauge fields  can be 
consistently taken
to zero we will not write 
$F_{\mu\nu}$
anymore.
$V(T)$ is the tachyon potential that
is symmetric under $T\rightarrow -T$
has maximum at $T=0$ 
equal to the 
tension of a non-BPS Dp-brane 
$\tau_p$
and has its minimum at $T=\pm \infty$
where it vanishes. 

Using the worldvolume
 diffeomorphism invariance 
we  can presume that the worldvolume
coordinates $\xi^{\mu}$ are equal 
to the spacetime coordinates $y^\mu$.
Explicitly, we can write
\begin{equation}
Y^{\mu}=\xi^{\mu} \ .
\end{equation}
Then the induced metric takes the form
\footnote{We restrict ourselves
in this paper to the situations
when  the 
background
metric  is diagonal.}
\begin{equation}
\gamma_{\mu\nu}\equiv
g_{MN}\partial_\mu X^M
\partial_\nu X^N=
g_{\mu\nu}+g_{mn}
\partial_\mu Y^m\partial_\nu Y^n
\ , 
\end{equation}
 where $Y^m \ , m,n=p+1,\dots,9$
parametrise the embedding of
Dp-brane in  a  space transverse
to its worldvolume. 
We should
also mention that generally 
the metric
components and dilaton are functions
of $\xi^\mu$ and $Y^m$:
\begin{equation}
g_{MN}=g_{MN}(\xi^\mu,Y^m) \ ,
\Phi=\Phi(\xi^\mu,Y^m) \ . 
\end{equation}
Now the equation of motion 
for 
$T$ and $Y^m$ that follow
from (\ref{acg}) take the form
\begin{equation}\label{eqtg}
\frac{\delta V}{\delta T}
e^{-\Phi}\sqrt{-\det \bA}
-\partial_\mu\left[e^{-\Phi}
V\sqrt{-\det \bA}\partial_\nu T 
\bAi^{\nu\mu}\right]=0 \
\end{equation}
and
\begin{eqnarray}\label{eqym}
\frac{\delta [e^{-\Phi}]}
{\delta Y^m}V\sqrt{-\det \bA}+
\frac{e^{-\Phi}}{2}V\left[
\frac{\delta g_{\mu\nu}}
{\delta Y^m}
+\frac{\delta g_{np}}
{\delta Y^m}\partial_\mu Y^n\partial_\nu Y^p\right]
\bAi^{\nu\mu}
\sqrt{-\det\bA}-
\nonumber \\
-\partial_\mu\left[
e^{-\Phi}Vg_{mn}
\partial_\nu Y^n
\bAi^{\nu\mu}
\sqrt{-\det\bA}\right]=0 \ . 
\nonumber \\
\end{eqnarray}
Our goal is to find
the solution of these equations of motions
that can be interpreted as a lower
dimensional D(p-1)-brane.
In order to obtain such a solution
we will closely follow
the paper by A. Sen 
\cite{Sen:2003tm}.
Let us choice one
particular worldvolume coordinate,
say $\xi^p\equiv x$ and
consider following ansatz
for the fields living
on the worldvolume of
Dp-brane
\begin{equation}\label{ans}
T=f(a(x-t(\xi))) \ ,
Y^m=Y^m(\xi) \ , 
\end{equation}
where  $\xi^\alpha \ ,
\alpha=0,\dots,p-1$ 
are coordinates
tangential to the kink worldvolume.
As in \cite{Sen:2003tm} we presume
that $f(u)$ satisfies following
properties
\begin{equation}
f(-u)=-f(u) \ ,
f'(u)>0 \ , \forall u \ ,
f(\pm \infty)=\pm \infty \ 
\end{equation}
but is otherwise an arbitrary 
function of its argument $u$. 
$a$ is a constant that we shall
take to $\infty$ in the end.
In this limit we have $T=\infty$ for
$x>t(\xi)$ and $T=-\infty$ 
for $x<t(\xi)$. 

Our goal is to check that the
ansatz (\ref{ans}) solves the
equation of motion 
(\ref{eqtg}) and (\ref{eqym}). 
Firstly, using
(\ref{ans})  the
matrix 
$\bA_{\mu\nu}$ takes the form
\begin{eqnarray}
\bA_{xx}=g_{xx}+a^2f'^2 \ ,
\bA_{x\alpha}=g_{x\alpha}-a^2f'^2
\partial_\alpha t \ , \nonumber \\
\bA_{\beta x}=g_{\beta x}-
a^2f'^2\partial_\beta t \ ,
\bA_{\alpha\beta}=(a^2f'^2-g_{xx})
\partial_\alpha t\partial_\beta t+
\mat_{\alpha\beta} \ ,
\nonumber \\
\mat_{\alpha\beta}=
g_{\alpha\beta}+g_{mn}
\partial_\alpha Y^m\partial_\beta
Y^n+\partial_\alpha t\partial_\beta t \ .
 \nonumber \\
\end{eqnarray}
For next purposes it will
be useful to know the form of the
determinant $\det\bA$.
Using the following identity
\begin{equation}
\det\bA=
\det (\bA_{\alpha\beta}-
\bA_{\alpha x}\frac{1}{\bA_{xx}}
\bA_{x\beta})
\det \bA_{xx}
\end{equation}
we get
\begin{equation}\label{dea}
\det\bA=
a^2f'^2\det(\mat_{\alpha\beta})
+O(1/a) \ .
\end{equation}
As a next step we should 
express $\bAi$ in terms of 
$\mat$. After some
calculations we find  
\begin{eqnarray}\label{ina}
\bAi^{xx}=(\mat^{-1})^{\alpha\beta}
\partial_\alpha t\partial_\beta t \ ,
\bAi^{x\beta}=
\partial_\alpha t (\mat^{-1})
^{\alpha \beta} \ ,
 \nonumber \\
\bAi^{\alpha x}=
(\mat^{-1})^{\alpha\beta}\partial_\beta t \ ,
\bAi^{\alpha\beta}=
(\mat^{-1})^{\alpha\beta}
 \  \nonumber \\
\end{eqnarray}
up to corrections of order 
$\frac{1}{a^2}$. 
In the following calculation
we will also need an 
exact relation
\begin{equation}
\bAi^{\mu x}-
\bAi^{\mu\alpha}\partial_\alpha t=
\frac{1}{a^2f'^2}
\left(\delta^{\mu}_x-
\bAi^{xx}g_{xx}\right) \ .
\end{equation}
Using this expression we can
now  write 
\begin{eqnarray}
\partial_\mu \left[e^{-\Phi}V
\sqrt{-\det\bA}
 \bAi^{\mu\nu}\partial_\nu T
\right]=\partial_\mu \left[
e^{-\Phi}Vaf'\frac{1}{a^2f'^2}
(\delta^{\mu}_x -
\bAi^{\mu x}g_{xx})\sqrt{-\det\bA}\right] \ .  
\nonumber \\
\end{eqnarray}
Following \cite{Sen:2003tm} we can
now argue that due to the explicit
factor of $a^2f'^2$ in the 
denominator
 the 
leading contribution
from individual terms in this expression is now
of order $a$ and hence we 
can use the approximative
results of $\det\bA$ and $\bAi$
given in  (\ref{dea}) and (\ref{ina})
 to analyse the equation of motion
for tachyon
\begin{eqnarray}
\partial_\mu \left[
e^{-\Phi}V\sqrt{-\det\bA}
af'\frac{1}{a^2f'^2}
(\delta^{\mu}_x -\bAi^{\mu x}g_{xx})\right]
-\nonumber \\
-e^{-\Phi}V'
\sqrt{-\det\bA}=
\nonumber \\
\partial_x\left[e^{-\Phi}V
\sqrt{-\det \mat}
(1-\mati^{\alpha\beta}g_{xx}
\partial_\alpha t
\partial_\beta t)\right]-
\nonumber \\
-\partial_\alpha
\left[e^{-\Phi}
V\sqrt{-\det\mat}
\mati^{\alpha\beta}g_{xx}
\partial_\beta t\right]
-af'e^{-\Phi}V'
\sqrt{-\det\mat}=
\nonumber \\
=V\left\{\partial_x
\left[e^{-\Phi}
\sqrt{-\det \mat}
(1-\mati^{\alpha\beta}g_{xx}\partial_\alpha t
\partial_\beta t
)\right]-
\right.\nonumber \\
-\left. \partial_\alpha
\left[e^{-\Phi}\sqrt{-\det \mat
}
(\mati^{\alpha\beta}
g_{xx}\partial_\beta t)\right] \right\}=0 \ .
\nonumber \\
\end{eqnarray}
This is important result that
deserves deeper explanation. 
Firstly,
from the form of the tachyon potential
in the limit $a\rightarrow \infty$ we
know that $V$ is equal to zero for
$x-t(\xi)\neq 0$ while for $
x-t(\xi)=0$ we get $V(0)=\tau_p$. 
Then it is clear that the tachyon
equation of motion is obeyed for
$x-t(\xi)\neq 0$ while 
for $x=t(\xi)$ we should demand that
 the expression in the
bracket should be equal to zero.
In other words, we obtain following
equation 
\begin{eqnarray}\label{eqtf}
\frac{\delta e^{-\Phi}}
{\delta x}
\sqrt{-\det \mat}
+\frac{e^{-\Phi}}{2}
\left(\frac{\delta g_{\alpha\beta}}
{\delta x}+
\frac{\delta g_{xx}}
{\delta x}\partial_\alpha t
\partial_\beta t+
\frac{\delta g_{mn}}
{\delta x}\partial_\alpha Y^m
\partial_\beta Y^n\right)
\mati^{\beta\alpha}
\sqrt{-\det\mat}-
\nonumber \\
-\partial_\alpha
\left[e^{-\Phi}
\sqrt{-\det \mat}
(\mati^{\alpha\beta}
g_{xx}\partial_\beta t)\right]
-\partial_x\left[e^{-\Phi}
\sqrt{-\det\mat}\mati^{\alpha
\beta}g_{xx}\right]\partial_\alpha t
\partial_\beta t=0 \ .
\nonumber \\
\end{eqnarray}
We must stress that in (\ref{eqtf})
we  firstly perform
the derivative with respect to
 $x$ and
then we insert the value $x=t(\xi)$
back to the resulting equation
of motion. 
For example, in the first term
on the second line we should perform
a derivative with 
respect to $\xi^\alpha$
with in mind that $x$ is an
independent variable. 
After doing  this  we should 
everywhere replace $x$ with
$t(\xi)$.  Then the presence of
the second term on the second
line is crucial for  an
interpretation
of $t(\xi)$ as an additional 
scalar field that parametrises the
position of D(p-1)-brane in $x$ 
direction.

Put differently,  we expect that the tachyon 
condensation
leads to an emergence of 
D(p-1)-brane
that is localised at $x=t(\xi)$.
For that reason we should
 compare the equation (\ref{eqtf})
  with the equation of motion
for D(p-1)-brane embedded in
the same background. 
As we know
the dynamics of such a Dp-brane
is governed by the DBI action
\begin{equation}\label{actDBI}
S=-T_{p-1}\int 
d^{p}\xi e^{-\Phi}
\sqrt{-\det \bA_{\alpha\beta}^{BPS}} \ , 
\end{equation}
where the matrix 
$\bA^{BPS}_{\alpha\beta}$
 takes the form
\begin{equation}\label{babps}
\bA^{BPS}_{\alpha\beta}=
g_{\alpha\beta}+
g_{xx}\partial_\alpha Y
\partial_\beta Y+
g_{mn}\partial_\alpha Y^m
\partial_\beta Y^n  \ , 
m,n=p+1,\dots,9 \ ,
\end{equation}
and where $T_{p-1}$ is the
tension of BPS D(p-1)-brane.
Recall that $T_p$ is 
is related to the tension 
of the non-BPS D(p-1)-brane 
$\tau_p$  
as $\tau_{p-1}=\sqrt{2}T_{p-1}$.
In (\ref{babps})  
we have
chosen one particular 
transverse mode
$Y$ in order to have a 
contact with the 
mode $t$ defined in the
equation (\ref{ans}).
Finally,
the scalar fields $Y^m$ 
have the same meaning as in
the case of a
 non-BPS Dp-brane.
Now the equations of motion that 
follow from 
(\ref{actDBI})
take the form
\begin{eqnarray}\label{Ymeq}
\frac{\delta }
{\delta Y^m}
\left[ e^{-\Phi}\sqrt{-
\det\bA^{BPS}_{\alpha\beta}}\right]
-\partial_{\alpha}
\left[e^{-\Phi}
\sqrt{-\det \bA^{BPS}_{\alpha\beta}}
\bAi_{BPS}^{\beta\alpha}
g_{mn}\partial_\beta Y^n
\right]=0 \ ,
\nonumber \\
\frac{\delta }
{\delta Y}
\left[ e^{-\Phi}\sqrt{-
\det\bA^{BPS}_{\alpha\beta}}\right]
-\partial_{\alpha}
\left[e^{-\Phi}
\sqrt{-\det \bA^{BPS}_{\alpha\beta}}
\bAi_{BPS}^{\beta\alpha}
g_{xx}\partial_\beta Y
\right]=0 \ ,
\nonumber \\
\end{eqnarray}
where the variation 
$\frac{\delta }{\delta Y^m}
\ , \frac{\delta }{\delta Y}$
means the variation of the metric, 
dilaton
 with respect to $Y^M, \ Y$
respectively.
 Explicitly,
the equation of motion for $Y$ 
can be written as
\begin{eqnarray}\label{Yeq}
\frac{\delta e^{-\Phi}}
{\delta Y}
\sqrt{-\det\bA}_{BPS}+
\nonumber \\
\frac{1}{2}e^{-\Phi}
\left(\frac{\delta g_{\alpha\beta}}
{\delta Y}+\frac{\delta g_{xx}}
{\delta Y}\partial_\alpha Y
\partial_\beta Y+
\frac{\delta g_{mn}}
{\delta Y}\partial_\alpha Y^m
\partial_\beta Y^n\right)
\bAi^{\beta\alpha}_{BPS}
\sqrt{-\det\bA}_{BPS}-
\nonumber \\
-\partial_{\alpha}
\left[e^{-\Phi}
\sqrt{-\det \bA}_{BPS}
\bAi_{BPS}^{\alpha\beta}g_{xx}
\partial_\beta Y
\right]=0 \ . 
\nonumber \\
\end{eqnarray}
To see more clearly the relation
with  the
equation (\ref{eqtf}) note
that the expression on the
third line can be written as
\begin{eqnarray}
\partial_{\alpha}
\left[e^{-\Phi}
\sqrt{-\det \bA}_{BPS}
\bAi_{BPS}^{\alpha\beta}g_{xx}
\partial_\beta Y
\right]=\nonumber \\
=\partial_\alpha \left[
e^{-\Phi(\xi,x)}
\sqrt{-\det\bA_{BPS}(\xi,x)}
\bAi^{\alpha\beta}_{BPS}(\xi,x)
\partial_\beta Y\right]
\nonumber \\
+\partial_x
\left[e^{-\Phi(\xi,x)}
\sqrt{-\det\bA_{BPS}(\xi,x)}
\bAi^{\beta\alpha}_{BPS}(\xi,x)
\right]
\partial_\alpha Y\partial_\beta Y \ ,
\nonumber \\
\end{eqnarray}
where on the second line the derivative
with respect to $\xi^\alpha$ 
treats $x$ as an independent variable
so that we firstly perform derivative with
respect to $\xi^\alpha$ and then 
we replace
$x$ with $Y$. We proceed in the
same way with  the expression 
on the third line where we firstly
perform the  variation with respect
to $x$ and  then we replace 
$x$ with $Y$.  
Now it is clear that this prescription is
the same as the expression on the
second line in  (\ref{eqtf}).   
More precisely,   if we compare
the equation 
(\ref{Yeq}) with the
the equation
(\ref{eqtf}) we see that these two
expressions coincide 
when we identify
$t$ with $Y$. In other words,
the location of the tachyon kink
is completely determined 
by field $t(\xi)$ that obeys the
equation of motion of the 
embedding
mode of D(p-1)-brane. 
We mean that
this is very satisfactory result
that shows 
that the Sen's construction
of the tachyon kink 
can be consistently performed
in curved background as well.

Now we will discuss 
the equation
of motion for $Y^k$. 
Again, we will proceed as
in \cite{Sen:2003tm}.
We begin with  the first term in 
(\ref{eqym})
that for the ansatz (\ref{ans})
takes the form
\begin{equation}\label{eqymp1}
\frac{\delta e^{-\Phi}}
{\delta Y^k}
V\sqrt{-\det\bA}=
af'V\frac{\delta e^{-\Phi}}
{\delta Y^k}\sqrt{-\det\mat} \ .
\end{equation}
In the same way we can 
show that
  the second term
in (\ref{eqym}) can be 
written
as
\begin{eqnarray}\label{eqymp2}
e^{-\Phi}V\left[
\frac{\delta g_{\mu\nu}}
{\delta Y^k}
+\frac{\delta g_{mn}}
{\delta Y^k}
\partial_\mu Y^m\partial_\nu Y^n\right]
\bAi^{\nu\mu}
\sqrt{-\det\bA}=
af'Ve^{-\Phi}\sqrt{-\det\mat}\times
\nonumber\\
\left[\frac{\delta g_{xx}}
{\delta Y^k}
\mati^{\alpha\beta}\partial_\alpha t
\partial_\beta t+(
\frac{\delta g_{\alpha\beta}}
{\delta Y^k}+
\frac{g_{mn}}{\delta Y^k}
\partial_\alpha Y^m\partial_\beta Y^n)
\mati^{\alpha\beta}\right] \ . 
\nonumber \\
\end{eqnarray}
Finally,  the third term
in (\ref{eqym}) is
equal to
\begin{eqnarray}\label{eqymp3}
-\partial_\mu\left[
e^{-\Phi}Vg_{km}\partial_\nu Y^m
\bAi^{\nu\mu}
\sqrt{-\det\bA}\right]=\nonumber \\
-\partial_x\left[Ve^{-\Phi}
g_{km}\partial_\alpha Y^m
\bAi^{\alpha x}\sqrt{-\det\bA}\right]
-\partial_\alpha
\left[Ve^{-\Phi}g_{km}
\partial_\beta Y^m
\bAi^{\alpha\beta}
\sqrt{-\det\bA}\right]=
\nonumber \\
=-af'\partial_x
\left[Ve^{-\Phi}g_{km}
\partial_\alpha Y^m
\mati^{\alpha\beta}
\partial_\beta t
\sqrt{-\det\mat}\right]-
\nonumber\\
-af'\partial_\alpha \left[Ve^{-\Phi}
g_{km}\partial_\beta Y^m
\mati^{\alpha\beta}
\sqrt{-\det\mat}
\right]=
\nonumber \\
-af'V\left(\partial_\alpha\left[e^{-\Phi}
g_{km}\partial_\beta Y^m
\mati^{\alpha \beta}
\sqrt{-\det\mat}\right]
+\partial_x\left[
e^{-\Phi}g_{km}\partial_\alpha Y^m
\mati^{\alpha\beta}\partial_\beta t
\sqrt{-\det\mat}\right]
\right) \nonumber \\
\end{eqnarray}
using the fact that 
$\partial_\alpha f=
-af'\partial_\alpha t$.
Then collecting 
(\ref{eqymp1}), (\ref{eqymp2}) and
(\ref{eqymp3}) together 
 we obtain
\begin{eqnarray}\label{Vym}
V\left\{
\frac{\delta e^{-\Phi}}
{\delta Y^k}
\sqrt{-\det\mat}
+\frac{1}{2}e^{-\Phi}\sqrt{-\det\mat}\times
\right.\nonumber\\
\left.\times\left[\frac{\delta g_{xx}}{\delta Y^k}
\partial_\alpha t
\partial_\beta t+
\frac{\delta g_{\alpha\beta}}
{\delta Y^k}+\frac{g_{mn}}{\delta Y^k}
\partial_\alpha 
Y^m\partial_\beta Y^n
\right] \mati^{\alpha\beta}\right.
\nonumber \\
\left.-\partial_\alpha\left[e^{-\Phi}
g_{km}\partial_\beta Y^m
\mati^{\alpha \beta}
\sqrt{-\det\mat}\right]
-\partial_x\left[
e^{-\Phi}g_{km}\partial_\alpha Y^m
\mati^{\alpha\beta}
\sqrt{-\det\mat}\right]\partial_\beta t\right\}=0 \ .
 \nonumber \\
\end{eqnarray}
Again it is easy to see that for
$x\neq t(\xi)$ 
the potential vanishes
for $a\rightarrow \infty$ while 
for
$x=t(\xi)$ we have $V(0)=\tau_p$.
Then  in order to obey the
equation of motion the 
expression  in the bracket
should be equal to zero for
$x=t(\xi)$. 
Note also that 
in the second expression
on the last line 
   in (\ref{Vym})
we firstly perform   a derivative
with respect to 
$x$ and then 
we replace $x$ with $t(\xi)$.
 In other words,  we can rewrite
 the last line 
in (\ref{Vym}) into the form 
\begin{equation}
\partial_\alpha
\left[e^{-\Phi(t(\xi))}
\sqrt{-\det\mat(t(\xi))}
\mati^{\alpha\beta}(t(\xi))
g_{km}(t(\xi))
\partial_\beta\beta Y^m\right] \ ,
\end{equation}
where we have explicitly stressed
the dependence of the action
on the mode $t(\xi)$ that replaces
in the action the dependence on
 $x$. 
This result again supports the
claim that we should  
identify
 $t(\xi)$ with
an additional embedding 
coordinate
of the D(p-1)-brane. 
Then by
comparing the expression in 
the bracket in (\ref{Vym}) 
with the equation
of motion for $Y^m$ given
in (\ref{Ymeq})
we see
that these two expressions
coincide.  
In summary, we have shown that
the tachyon kink solution on
a non-BPS Dp-brane in nontrivial
background can be identified as
a lower dimensional D(p-1)-brane
that is localised at the core
of the kink. We have also shown
that the dynamics of this
D(p-1)-brane is governed by
DBI action.
\subsection{Stress energy tensor}
Further support for the 
interpretation
of the tachyon kink as 
a lower dimensional
D(p-1)-brane 
can be obtained 
 from 
the analysis
of the stress energy tensor
for the non-BPS Dp-brane.
In order to find its form recall 
that   we 
can write the action (\ref{acg})
as 
\begin{equation}\label{dactem}
S_{p}=-\int d^{10}yd^{(p+1)}
\xi\delta
(Y^M(\xi)-y^M)e^{-\Phi}V(T)
\sqrt{-\det \bA} \ ,  
\end{equation}
where
\begin{equation}
\bA_{\mu\nu}
=G_{MN}\partial_\mu 
Y^M\partial_{\nu}Y^N+
\partial_{\mu}T
\partial_{\nu}T \ ,
\end{equation}
and where $\xi^\mu \ , \mu=0,\dots,p$
are worldvolume 
coordinates on Dp-brane.
The form of
action (\ref{dactem})
is  useful for
determining the stress energy 
tensor $T_{MN}(y)$ of an unstable
D-brane. In fact, the stress
energy tensor $T_{MN}(y)$  is defined 
as the variation of $S_p$
with   
respect
to $g_{MN}(y)$ 
\begin{eqnarray}\label{TMNg}
T_{MN}(y)=-2
\frac{\delta S_{p}}{
\sqrt{-g(y)}\delta g^{MN}(y)}=\nonumber \\
=-\int d^{(p+1)}\xi\frac{\delta(Y^M(\xi)
-y^M)}
{\sqrt{-g(y)}}e^{-\Phi}V
g_{MK}g_{NL}
\partial_{\mu}Y^K\partial_{\nu}
Y^L(\bA^{-1})^{\nu\mu}
\sqrt{-\det \bA} \ . \nonumber \\ 
\end{eqnarray}
The form of the stress energy
tensor for gauge fixed 
Dp-brane action  can be obtained
from   (\ref{TMNg}) by imposing
the condition
\begin{equation}
Y^{\mu}=\xi^{\mu} \ , 
\mu=0,1,\dots,p \ .
\end{equation}
Then the integration over $\xi^{\mu}$
swallows up the delta function
$\delta (y^{\mu}-Y^{\mu}(\xi))=
\delta(y^{\mu}-\xi^{\mu})$ so that
the resulting stress energy tensor
takes the form 
\begin{eqnarray}
T_{mn}=-\frac{\delta(Y^m(\xi)-
y^m)}{\sqrt{-g}}
e^{-\Phi}Vg_{mm}\partial_\mu
Y^mg_{nn}\partial_\nu Y^n
\bAi^{\nu\mu}\sqrt{-\det\bA} \ ,
\nonumber \\
T_{\mu\nu}=
-\frac{\delta(Y^m(\xi)-y^m)}
{\sqrt{-g}}e^{-\Phi}Vg_{\mu\mu}
g_{\nu\nu}\bAi^{\nu\mu}
\sqrt{-\det\bA} \ ,
\nonumber \\ 
T_{\mu n}=
-\frac{\delta(Y^m(\xi)-
y^m)}
{\sqrt{-g}}e^{-\Phi}Vg_{\mu\mu}
g_{nn}\partial_\nu Y^n
\bAi^{\nu\mu}\sqrt{-\det\bA} \ ,
\nonumber \\ 
T_{m\nu}
=-\frac{\delta(Y^m(\xi)-y^m)}
{\sqrt{-g}}e^{-\Phi}V
g_{mm}\partial_\mu 
Y^mg_{\nu\nu}\bAi^{\nu\mu}
\sqrt{-\det\bA} \ ,
\nonumber \\
\end{eqnarray}
using
 the fact that the metric 
is diagonal.

If we  now insert the ansatz
(\ref{ans}) into these expressions
we get
\begin{eqnarray}\label{Tans}
T_{mn}=-\frac{\delta(Y^m(\xi)-x^m)}
{\sqrt{-g}}
Vaf'e^{-\Phi}g_{mm}\partial_\alpha
Y^mg_{nn}\partial
_\beta Y^n
\mati^{\alpha\beta}\sqrt{-\det\mat} \ ,\nonumber \\
T_{\alpha\beta}=-
\frac{\delta(Y^m(\xi)-y^m)}{\sqrt{-g}}
Vf'ae^{-\Phi}
g_{\alpha\alpha}
g_{\beta\beta}\mati^{\alpha\beta}
\sqrt{-\det\mat} \ , \nonumber \\
T_{xx}=
-\frac{\delta(Y^m-y^m)}{\sqrt{-g}}
Vf'ae^{-\Phi}
g_{xx}\partial_\alpha t\partial_\beta t
\mati^{\alpha\beta}\sqrt{-\det\mat} \ ,
\nonumber \\
T_{x\alpha}=T_{\alpha x}=0 \ ,
\nonumber \\
T_{mx}=T_{xm}=-
\frac{\delta (Y^m-y^m)}{
\sqrt{-g}}Vf'ae^{-\Phi}g_{mm}
\partial_\alpha Y^m g_{xx} 
\partial_\beta 
t\mati^{\alpha\beta}\sqrt{-\det\mat} \ .
\nonumber \\
\end{eqnarray}
From now on the 
notation $Y^m(\xi),
t(\xi)$ means that
 these fields are
functions of the coordinates on
the worldvolume of the kink $\xi^\alpha,
\alpha=0,\dots,p-1$. 

According to \cite{Sen:2003tm}
the components of the stress energy
tensor of the lower dimensional D(p-1)-brane
arise by integrating 
all $T_{MN}$ given above 
over the direction
of the tachyon condensation
that in our case is  $x$.
Now we should be more careful
since  metric components
generally  depend
on $x$.  Let us introduce the
following notation for the
components of the stress energy
tensors (\ref{Tans})
\begin{equation}
T_{MN}=V(f(a(t(\xi)-x)))af'
\tilde{T}_{MN}(x) \ ,
\end{equation}where we have 
explicitly 
stressed the dependence of
$\tilde{T}_{MN}$ on $x$. 
If we 
now integrate $T_{MN}$ over
$x$ we get  
\begin{equation}
T_{MN}^{kink}=
\int_\infty^\infty
 dx V(f(a(x-t(\xi))))f'a
\tilde{T}_{MN}(x)=
\int dm V(m)\tilde{T}_{MN}\left(
\frac{f^{-1}(m)}{a}+
t(\xi)\right) \ .
\end{equation}
 In the limit $a\rightarrow 
\infty$ the term proportional
to $1/a$ goes 
to zero and we
get that the components 
$\tilde{T}_{MN}$ are functions
of $t(\xi)$ in place of $x$.
Further, we can argue,
following \cite{Sen:2003tm}
that the exponential
fall off in $V(m)$ implies
that in the limit
 $a\rightarrow
\infty$ the contribution to
the stress energy
 tensor is
localised at the point where 
$V$ is equal to $V(0)=\tau_{p}$
which happens for $x=t(\xi)$.
In other words, when we presume that
the tension of BPS D(p-1)-brane
is given by the integral
\begin{equation}
T_{p-1}=
\int_{-\infty}^{\infty}dm V(m)
\end{equation}
we obtain the result that
the components of the stress
energy tensor of the kink
take the form \begin{eqnarray}\label{Tansf}
T^{kink}_{mn}=-
\frac{T_{p-1}\delta(Y^m(\xi)-y^m)
\delta(t(\xi)-x)}
{\sqrt{-g}}
e^{-\Phi}g_{mm}\partial_\alpha
Y^mg_{nn}\partial_\beta 
Y^n
\mati^{\alpha\beta}
\sqrt{-\det\mat} \ ,
\nonumber \\
T^{kink}_{\alpha\beta}=-
\frac{T_{p-1}\delta(Y^m(\xi)-y^m)
\delta(t(\xi)-x)}{\sqrt{-g}}
e^{-\Phi}g_{\alpha\alpha}
g_{\beta\beta}\mati^{\alpha\beta}
\sqrt{-\det\mat} \ , \nonumber \\
 T^{kink}_{xx}=
-\frac{T_{p-1}\delta(Y^m-y^m)
\delta(t(\xi)-x)}{\sqrt{-g}}
e^{-\Phi}
g_{xx}\partial_\alpha t\partial_\beta t
\mati^{\alpha\beta}
\sqrt{-\det\mat} \ ,
\nonumber \\
T_{x\alpha}^{kink}=T^{kink}_{\alpha x}=0 \ ,
\nonumber \\
T_{mx}^{kink}=
T_{xm}^{kink}=-
\frac{T_{p-1}\delta (Y^m-y^m)
\delta(t(\xi)-x)}{
\sqrt{-g}}e^{-\Phi}g_{mm}
\partial_\alpha Y^m g_{xx} 
\partial_\beta 
t\mati^{\alpha\beta}
\sqrt{-\det\mat} \ ,
\nonumber \\ 
\end{eqnarray}
where  it is understood
that $g_{MN}$ and 
$\Phi$ are
 functions of
$\xi^{\alpha} \ , Y^m(\xi) \ ,
 t(\xi)$. In other words, 
the components of the 
stress energy tensors (\ref{Tansf}) 
correspond to the components
of the stress energy tensor 
of  a D(p-1)-brane
localised at the points
$Y^m(\xi),t(\xi)$.
\section{Examples of the tachyon
condensation on a non-BPS Dp-brane
in nontrivial   background}
\label{third}
In this section we will briefly
discuss  some
examples of the tachyon condensation on
a non-BPS Dp-brane that is embedded
in  nontrivial  backgrounds.
\subsection{NS5-brane background}
As the first example  we
will consider the        
 background  corresponding
to the stack of $N$ coincident 
NS5-branes
\begin{eqnarray}\label{NSbac}
ds^2=dx_{\mu}dx^{\mu}+H_{NS}
dx^mdx^m \ , \nonumber \\
e^{2\Phi}=
H_{NS} \ , 
\nonumber \\
H_{mnp}=-\epsilon^q_{mnp}
\partial_q
\Phi \ ,  \nonumber \\
\end{eqnarray}
where the harmonic function
$H_{NS}$ for $N$ coincident
NS5-branes is equal to
\begin{equation}
H_{NS}(y^m)=1+\frac{2\pi N}{y^my_m} \ ,
\end{equation}
where $y^m \ , m=6,\dots,9$ 
label directions transverse to the
worldvolume of NS5-branes.

The most simple case occurs 
when Dp-brane
is stretched in the 
direction parallel
with the worldvolume of  NS5-branes
\footnote{In what follows we will consider
the situation when we can ignore
the NS two form background.}.
Using (\ref{NSbac})
it is then easy to determine the
worldvolume metric 
\begin{eqnarray}
g_{\mu\nu}=
\eta_{\mu\nu} \ ,
g_{m_1n_1}=\delta_{m_1n_1} \ , 
m_1 \ , n_1
=p+1,\dots,5 \ , \nonumber \\
g_{m_2n_2}=H_{NS}\delta_{m_2n_2} \ ,
m_2,n_2=6,\dots,9 \ ,
\nonumber \\
\end{eqnarray}
where now $H_{NS}$ is function
of $Y^{m_2}$
\begin{equation}
H_{NS}=1+\frac{2\pi N}{Y^{m_2}Y_{m_2}} \ .
\end{equation}
Thanks 
to the manifest 
$SO(p)$ symmetry
of the worldvolume theory 
all spatial coordinates
$\xi^i \ , i=1,\dots,p$ 
are equivalent. Then we choose
the direction on which the
tachyon depends to be $\xi^p=x$.
Now it is clear that the
spatial dependent tachyon condensation
studied in previous section
leads to the emergence of
a  D(p-1)-brane that is stretched in
the $x^0,\dots,x^{p-1}$ 
directions and
and which transverse position
is determined by the worldvolume
fields 
$t(\xi),Y^{m}(\xi)$. These fields
also  obey the
equations of motion that arise
from the DBI action for a
 D(p-1)-brane that
moves in the background of 
$N$ NS5-branes.

Another possibility occurs 
when we  consider   a non-BPS 
Dp-brane stretched in
 some of the 
transverse directions to the
worldvolume of NS5-branes.
More precisely, 
let us consider an unstable
Dp-brane
that is stretched in 
$x^0,x^1,\dots,x^k$ directions
and in $x^6,\dots,x^{6+p-k}$
directions.
Then the 
 metric components that appear
on the worldvolume of
the Dp-brane take the form 
\begin{eqnarray}
g_{\mu_1\nu_1}=
\eta_{\mu_1\nu_1} \ ,
\mu_1\ , \nu_1=0,\dots,k \ ,
\nonumber \\
g_{\mu_2\nu_2}=H_{NS}
\delta_{\mu_2\nu_2} \ ,
\mu_2 \ , \nu_2=6,\dots, (6+p-k) \ ,
\nonumber \\
g_{m_1n_1}=
\delta_{m_1n_1} \ ,
m_1 , n_1=k+1,\dots,5 \ , 
\nonumber \\
g_{m_2n_2}=H_{NS}\delta_{m_2 n_2} \ ,
m_2 \ , n_2=(7+p-k)\ , 
\dots, 9 \ ,
\nonumber \\
\end{eqnarray}
where the function $H_{NS}$ has
the form
\begin{equation}
H_{NS}=1+\frac{2\pi N}{(\xi^{\mu_2}
\xi_{\mu_2}+Y^{m_2}Y_{m_2})^2} \ .
\end{equation}
Now there are many
 possibilities how
to construct lower dimensional
 D(p-1)-brane. If we
perform the spatial dependent
 tachyon condensation on
the worldvolume of the
 non-BPS Dp-brane where
the tachyon $t(x)$ depends on 
coordinate from the set
$\xi^1,\dots,\xi^k$ (again, we
take $x=\xi^k$) we 
obtain  D(p-1)-brane
that is localised in $x^k$ 
direction
and that is stretched in $x^0,\dots,x^{k-1}$
and $x^{6},\dots, x^{(6+p-k)}$
directions. It is important 
to stress
that the resulting configuration
of $N$ NS5-brane and BPS D(p-1)-brane
is not in general stable. 
Rather the
dynamics of the BPS D(p-1)-brane
in the background of $N$ NS5-branes
is governed the 
equation of motions
(\ref{Yeq}). To find  stable configuration
we should perform the same analysis as
 in \cite{Tseytlin:1996hi}.

Another possibility is to 
consider the tachyon 
condensation
in direction from the set
$\xi^6,\dots, \xi^{6+p-k}$. Let us choose
$x\equiv \xi^6$. 
Then it is clear that the
tachyon condensation leads to the
emergence of D(p-1)-brane 
stretched
in $(x^0,\dots,x^k,x^7,\dots,x^{6+p-k})$
directions and where the scalar
fields on its worldvolume 
$t(\xi), Y^{m_1}\ , Y^{m_2}$
describing embedding of this D(p-1)-brane in
nontrivial background, obey the equations
of motions that arise from DBI action for
BPS D(p-1)-brane. 
\subsection{Non-BPS Dp-brane in 
Dk-brane background}
The second example that we will
consider in this paper, is the
 spatial dependent tachyon
condensation on  the worldvolume
of a non-BPS Dp-brane that moves
in the background of 
$N$ BPS Dk-branes. This
background is characterised by 
following metric
 and dilaton in the form
\begin{eqnarray}
ds^2=H_k^{-1/2}\eta_{\alpha\beta}
dx^\alpha dx^\beta +
H_k^{1/2}\delta_{mn}dx^mdx^n \ ,
 \nonumber \\
\alpha,\beta=0,\dots,k \ , m,n=k+1,\dots,9
 \nonumber \\
e^{-2\Phi}=H_k^{\frac{k-3}{2}} \ ,
\nonumber \\
\end{eqnarray}
where the harmonic function $H_p$ 
takes the form
\begin{equation}
H_k=1+\frac{Ng_s(2\pi)^{\frac{7-k}{2}}}{(y^my_m)^{\frac{7-k}{2}}} \ ,
\end{equation}
where $y^m \ , m=k+1,\dots,9$ label
the directions transverse to the worldvolume
of $N$ Dk--branes.

There is again many 
possibilities 
how to put in  a non-BPS Dp-brane in
this background. 
As the first possibility 
let us consider a  non-BPS Dp-brane
that is stretched in $x^0,\dots,x^p$ directions
and that is localised in $Y^{m_1}, m_1=p+1,\dots,k$ 
directions (parallel with the  worldvolume of 
Dk-branes). This Dp-brane is
 also localised in $Y^{m_2} \ ,
 m_2=k+1,\dots,9$ 
directions transverse 
to Dk-branes worldvolume.
 Now the 
 metric components on
its  worldvolume  take the form
\begin{equation}
g_{\mu\nu}=H_k^{-1/2}
\eta_{\mu\nu} \ ,
g_{m_1n_1}=H_k^{-1/2}\delta_{m_1n_1} \ ,
g_{m_2n_2}=H_k^{1/2}\delta_{m_2n_2} \ ,
\end{equation}
where $H_k$ depends on  
$Y^{m_2}Y_{m_2}$. It is clear
that the spatial dependent
tachyon condensation 
(Let us choose $x$ that appears in
the ansatz (\ref{ans}) to be
equal to $\xi^p$.)
 leads to an emergence of a 
D(p-1)-brane with
 the  worldvolume fields 
$Y^{m_1} \ , Y^{m_2}$ as well as
with the mode $t(\xi)$ that
parametrises the location
of D(p-1)-brane in $x^p$ direction.

Another possibility  
occurs when we 
consider Dp-brane 
where some  of its worldvolume
directions are parallel with
 the worldvolume of Dk-branes
and other ones are stretched in the 
directions transverse to 
Dk-brane. This situation can
be described by following
induced metric on the worldvolume
of non-BPS Dp-brane:
\begin{eqnarray}
g_{\mu_1\nu_1}=
H_k^{-1/2}\eta_{\mu_1\nu_1} \ , 
\mu_1\ ,
\nu_1=0,\dots, l \ ,
\nonumber \\
g_{\mu_2\nu_2}=H_{NS}^{1/2}
\delta_{\mu_2\nu_2} \ , \mu_2 \ , 
\nu_2=k+1,\dots, (k+1+p-l)
\ ,
\nonumber \\
g_{m_1n_1}=H_k^{-1/2}
\delta_{m_1n_1} \ , m_1 ,
n_1=l+1,\dots,k \ , \nonumber \\
g_{m_2n_2}=H_{NS}^{1/2}
\delta_{m_2 n_2} \ ,
m_2\ , n_2=(k+2+p-l)\ , \dots, 9 \ , 
\nonumber \\
\end{eqnarray}
where  the function $H_k$  
is equal to
\begin{equation}
H_k=1+\frac{Ng_s(2\pi)^{\frac{7-k}{2}}}
{(\xi^{\mu_2}\xi_{\mu_2}+
Y^{m_2}Y_{m_2})^{\frac{7-k}{2}}} \ .
\end{equation}
If now the tachyon depends on
one of the coordinates from
the set $\xi^1,\dots,\xi^l$, say $x=\xi^l$, 
  we obtain
a  D(p-1)-brane that is
 localised in 
$x^l$ direction and 
that is stretched in 
$x^0,\dots,x^{l-1}$ and $x^{k+1},
\dots, x^{(k+1+p-l)}$ directions.

The next possibility corresponds
 to the tachyon condensation
in the direction transverse 
to Dk-branes, say $\xi^{k+1}\equiv x$. 
Following the general recipe given
in previous section it is clear
that  this spatial dependent
tachyon condensation leads to 
an emergence of a D(p-1)-brane 
that is stretched in 
$(x^0,\dots,x^l,x^{k+2},\dots,
x^{k+2+p-l})$ directions and
its positions in the transverse 
space are described with the
worldvolume scalar fields 
$Y^{m_1}(\xi),Y^{m_2}(\xi)$ 
and also with
$t(\xi)$ that parametrises the
position of D(p-1)-brane in 
$x^{k+1}$ direction. It is also clear
that  these modes  obey the
equations of motions that follow
from the DBI action for probe
D(p-1)-brane in the background
of $N$ Dk-branes. 
Note also that 
the resulting configuration
of $N$ Dk-branes and 
D(p-1)-brane
is not generally stable
 \cite{Tseytlin:1996hi}.
\section{Conclusion}\label{fourth}
This paper was devoted to the study of the 
spatial dependent tachyon condensation on the
worldvolume of a non-BPS Dp-brane that is
moving in nontrivial background. We have shown
that this tachyon condensation leads to an
emergence of the D(p-1)-brane that is moving
in the same background and where the scalar
mode that determines the location of
the kink on a non-BPS Dp-brane worldvolume
 can be interpreted as 
a mode that 
 describes the transverse position of
D(p-1)-brane and that 
obeys the equation of motion that
follows from  DBI action for D(p-1)-brane.
We hope that this result is  a nice example
of the efficiently of 
the effective field theory description 
of the tachyon condensation 
and it also gives strong support for the form
of the Dirac-Born-Infeld form of the tachyon
effective action (\ref{acg}).

The extension of this paper is obvious. First
of all we would like to study the tachyon 
condensation when we take into account
nontrivial NS $B$ field and also nontrivial 
Ramond-Ramond field. It would be also interesting
to study the tachyon condensation on the
supersymmetric form of the non-BPS Dp-brane
action.  We hope to return to these 
problems in future. 

{\bf 
Acknowledgement}

This work was supported by the
Czech Ministry of Education under Contract No.
MSM 0021622409.


\end{document}